\documentclass[pdflatex,sn-mathphys-num]{sn-jnl}

\usepackage{graphicx}%
\usepackage{multirow}%
\usepackage{amsmath,amssymb,amsfonts}%
\usepackage{amsthm}%
\usepackage{mathrsfs}%
\usepackage[title]{appendix}%
\usepackage{xcolor}%
\usepackage{textcomp}%
\usepackage{manyfoot}%
\usepackage{booktabs}%
\usepackage{algorithm}%
\usepackage{algorithmicx}%
\usepackage{algpseudocode}%
\usepackage{listings}%

\usepackage{tabularx}

\theoremstyle{thmstyleone}%
\theoremstyle{thmstyletwo}%

\theoremstyle{thmstylethree}%

\begin{document}


\title[WSSSMB]{Weak-Strong Steady-State Microbunching Accelerator Light Source}


\author*[1]{\fnm{Xiujie} \sur{Deng}}\email{dengxiujie@mail.tsinghua.edu.cn}

\author*[1]{\fnm{Alexander Wu} \sur{Chao}}\email{chaoalexander5@gmail.com}

\author[2]{\fnm{Wenhui} \sur{Huang}}

\author*[2]{\fnm{Zhilong} \sur{Pan}}\email{panzl@mail.tsinghua.edu.cn}

\author*[2]{\fnm{Chuanxiang} \sur{Tang}}\email{tang.xuh@tsinghua.edu.cn}

\author[2]{\fnm{Jingyuan} \sur{Zhao}}

\affil[1]{\orgdiv{Institute for Advanced Study}, \orgname{Tsinghua University},\\ \orgaddress{\city{Beijing 100084},\country{China}}}

\affil[2]{\orgdiv{Department of Engineering Physics}, \orgname{Tsinghua University},\\ \orgaddress{\city{Beijing~100084}, \country{China}}}

\abstract{We propose a phase space manipulation involving one energy modulation sandwiched by two dispersion sections which converts a bunched particle beam or bunch train to ultra-high-harmonic density modulation, while the energy modulation in principle can be arbitrarily weak. The same scheme can also be used for energy bunching, creating energy levels in a bunched beam. We further propose a mechanism invoking three laser modulators in a storage ring to longitudinally focus the electron beam both weakly and strongly, such that a microbunch train and its high-density-harmonics or energy bunching form and sustain turn-by-turn. We call this mechanism weak-strong steady-state microbunching (Weak-Strong SSMB). The longitudinal beta function can vary by seven orders of magnitude along such a ring, with the minimal value squeezed to 10 nm. An example application of Weak-Strong SSMB for kW coherent EUV radiation is presented. Extension to X-ray can be anticipated. An energy-leveled electron beam enables $\gamma$-ray frequency comb production. The ideas can be scaled to wavelengths like RF and THz, for bunch length and energy spread control, ultrashort X-ray and coherent THz generation. Our work establishes a new paradigm for longitudinal dynamics study, accelerator light source development, and opens great potential for accelerator physics and technology.}

\keywords{bunched beam harmonic generation, energy bunching,  weak-strong focusing, steady-state microbunching,  coherent radiation, extreme ultraviolet, energy spread and bunch length control, $\gamma$-ray frequency comb, high-luminosity collider}



\maketitle

\section{Introduction}\label{sec1}

Among the four fundamental forces, electromagnetism is the most rigorously understood and widely exploited. This is exemplified by accelerator-based light sources, which represent a pinnacle of our ability to harness electromagnetic interactions. When charged particle is accelerated, it radiates out electromagnetic wave. Electron as the lightest charged particle serves as a natural emitter. Once the prescribed motion of an electron is given, we can obtain the radiation characteristics. For an electron beam, the radiation property is determined by both the reference particle prescribed motion and also the 6D phase space distribution of the beam, or equivalently the prescribed motion of all the electrons in the beam. 

The situation becomes more involved and rich when we consider the radiation reaction back on the electrons. For example, the equilibrium beam parameters in a storage ring-based synchrotron radiation source (SR) are determined by the balance between radiation damping and quantum excitation, which originate from the classical radiation reaction and the quantum fluctuation of photon emission, respectively. The shaping of the electron trajectory--from bending magnets to undulators as radiators--and the minimization of the equilibrium transverse emittance via innovative magnetic lattice designs--from FODO to double-bend, and then to multi-bend achromats--have improved the radiation brightness by orders of magnitude, with the transverse coherence having reached the diffraction limit in the X-ray range.

Having considered the impact on equilibrium beam parameters, the prescribed motion used in calculating the radiation in a SR is still unperturbed. If we make the electron beam and radiation co-propagate and continuously interact with each other, for example in a long undulator, then the prescribed motion can also be affected. A positive feedback loop may form and lead to the formation of microbunching. This is how a free-electron laser (FEL) works~\cite{Kondratenko1980Generating,Bonifacio1984Collective}.  Microbunching enables strong coherent radiation generation, whose power scales quadratically with the number of electrons, in contrast to the usual incoherent radiation whose dependence is linear.  Linac-based FEL at present can deliver X-ray laser with high peak power.

While ring-based SRs and linac-based FELs are serving the present cutting-edge applications, one important open question is what could a future accelerator light source be like? A promising direction in this context is the steady-state microbunching (SSMB) mechanism~\cite{Ratner2010SSMB,Ratner2011Reversible,Jiao2011Terahertz,Chao2016SSMB}, which combines the aforementioned points of light source development. In particular, SSMB uses laser modulator in a storage ring to longitudinally focus the electron beam, while in a conventional ring this is done by the RF cavity. With dedicated lattice, microbunching can form and sustain turn-by-turn in such a ring, enabling both high-power and high-repetition-rate coherent radiation generation.  Recently, SSMB research has seen significant advances, including the successful proof-of-principle experiment~\cite{deng2021experimental,kruschinski2024confirming}, the development of SSMB working scenarios and related concepts, and beam dynamics studies~\cite{Deng2020Single,Deng2021Courant,Zhang2021Ultra,Pan2026lattice,Li2023GLSF,Deng2026TLCSSMB,Deng2025Reversible,Deng2025Stochastic,Khan2017Ultrashort,Jiang2022synchrotron,Lu2025Lattice,Pan2025IBS,Bian2025Threshold,Tang2026Resistive,Zhao2025Method,Tsai2022Theoretical,Deng2024Springer,Tang2026Physics,Dai2026Longitudinal}. 

A core goal--and advantage--of SSMB is to produce high-power radiation at short wavelengths, such as EUV and X-ray. To generate coherent radiation at such short wavelength, the microbunch length should reach nm level. In the meanwhile, for high average power output, the required modulation laser power should be sufficiently low to allow a high-duty-cycle or continuous-wave operation of the modulation system. With the aim of developing an SSMB-based high-power short-wavelength source, we have carried out a systematic study of the longitudinal weak focusing (LWF), longitudinal strong focusing (LSF) and generalized longitudinal strong focusing (GLSF) SSMB scenarios~\cite{Deng2020Single,Deng2021Courant,Zhang2021Ultra,Pan2026lattice,Li2023GLSF,Pan2025IBS, Deng2026TLCSSMB,Tang2026Physics}. A short summary of the investigation is in order. The LWF SSMB, constrained by the practical lower limit of the storage ring phase slippage factor, can generate microbunches as short as $50\sim100$~nm. The LSF SSMB can produce nm-scale microbunches, but the required modulation laser power is on the order of 1 GW. For reference, the maximum stored laser power in an optical enhancement cavity is currently about 1 MW~\cite{Lu2025OEC}. For the GLSF SSMB, the optimization of intrabeam scattering and nonlinear transverse-longitudinal coupling dynamics is challenging. We recognize these difficulties are greatly alleviated when targeting longer radiation wavelengths.

Therefore, realizing high-power, short-wavelength coherent light sources in storage rings necessitates further breakthroughs. In this paper, we tackle this challenge by proposing a novel phase space manipulation method, BBHG, and a storage ring working mechanism, Weak-Strong SSMB, and present a solution for a kW-level coherent EUV source based on the latter.



\section{Results}

\subsection{BBHG Phase Space Manipulation Method}

The first key we need is an efficient microbunching scheme. Since it is difficult to directly generate an nm-scale bunch with a moderate laser power, we seek frequency up-conversion of the density modulation via the laser-induced energy modulation. Representative examples in this direction include high-gain harmonic generation (HGHG)~\cite{Yu1991} and echo-enabled harmonic generation (EEHG)~\cite{Stupakov2009Echo}, which employ one and two stages of ``modulation + dispersion section ($R_{56}$)", respectively. To achieve bunching at the 
$n$-th laser harmonic, HGHG requires an energy modulation strength of roughly 
$n$ times the beam energy spread. Even for EEHG, the first-stage modulation strength still needs to be about 1 to 2 times the natural energy spread to produce significant bunching at high harmonics. For an electron beam in a storage ring, the equilibrium energy spread is typically at $1\times 10^{-3}$ level and is relatively large. 

\begin{figure}[tb]
	\centering
	\includegraphics[width=0.85\linewidth]{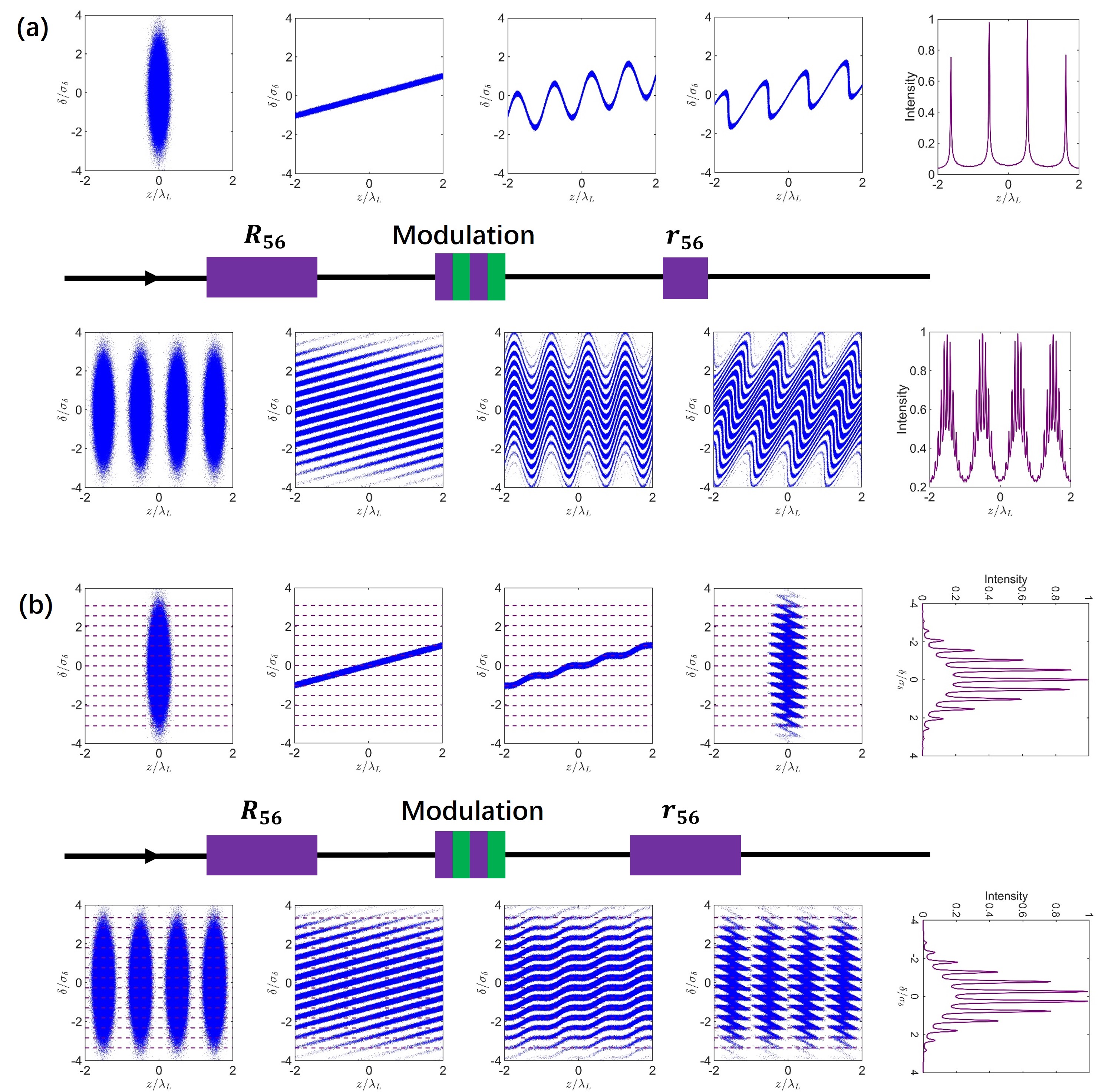}
	\caption{Schematic layout of the proposed BBHG phase space manipulation method, and the longitudinal phase space distribution evolution of a single bunch (top) or a bunch train (bottom) in such a scheme. (a) is for density bunching, (b) is for energy bunching. }
	\label{fig:BBHG}
\end{figure}

Here we propose a novel phase space manipulation method as shown in Fig.~\ref{fig:BBHG}a that involves one energy modulation sandwiched by two dispersion sections ($R_{56}$ and $r_{56}$) that can convert a bunched particle beam or bunch train to extremely-high-harmonic density modulation. We call such a scheme BBHG, which means  bunched beam or bunch train-based harmonic generation. When $|k_{\text{L}} R_{56}\sigma_{\delta \text{B}}|\gg1$ and $R_{56}/r_{56}=n$, the $n$-th laser harmonic bunching factor in BBHG is (see Methods)
\begin{equation}
b_{n}\equiv\langle e^{-ink_{\text{L}}z}\rangle\approx J_{n+1}(-nk_{\text{L}}r_{56}A_{\text{L}})\text{exp}\left[-{\left(k_{\text{L}}\sigma_{z\text{B}}\right)^{2}}/{2}\right],
\end{equation}
where $\langle\rangle$ means ensemble average, $k_{\text{L}}$ is the laser wavenumber, $J_{n+1}$ means the $n+1$-th order Bessel function of the first kind, $\sigma_{\delta\text{B}}$ and $\sigma_{z\text{B}}$ are the initial energy spread and bunch length, respectively, $A_{\text{L}}=eV_{\text{L}}/E_{0}$ is the laser imprinted energy modulation strength with $e$ the elementary charge, $V_{\text{L}}$ the laser modulation voltage and $E_{0}$ the electron energy. Note that this bunching factor is independent of the initial energy spread $\sigma_{\delta\text{B}}$. Also the exponential term does not depend on the harmonic number $n$. For $n>4$, let
$
|nk_{\text{L}}r_{56}A_{\text{L}}|=n+1+0.81(n+1)^{1/3},
$ we have the maximal bunching, 
$
b_{n}\approx\frac{0.67}{(n+1)^{1/3}}\text{exp}\left[-{(k_{\text{L}}\sigma_{z\text{B}})^2}/{2}\right]. 
$ The required laser modulation strength $A_{\text{L}}$ can be flexibly lowered by implementing a larger $r_{56}$. Since the manipulation is performed at the laser wavelength scale, the required absolute values of $R_{56}$ and  $r_{56}$ are actually quite modest, and their practical implementation is straightforward.

If we can make $k_{\text{L}}\sigma_{z\text{B}}\lesssim1$, the bunching factor can approach the theoretical maximal ${0.67}/{(n+1)^{1/3}}$.  For $\lambda_{\text{L}}=1030$ nm, this corresponds to $\sigma_{z\text{B}}\lesssim164$~nm, which is within the reach of a LWF SSMB ring.  Even if $\sigma_{z\text{B}}$ is twice this value, there is still respectful bunching at high harmonics. So by applying this BBHG method, we have solved the first key issue of an efficient microbunching generation technique. Motivated by ring-based coherent light source as it, this BBHG method can also be applied in a single-pass machine. It can also be scaled to other wavelength ranges like THz for coherent THz generation.

The method can also be used for energy bunching, creating energy levels in a bunched beam, as shown in Fig.~\ref{fig:BBHG}b. Such an energy-leveled  or energy distribution-shaped beam in general may find interesting applications in science. For example it can create $\gamma$-ray frequency comb when colliding with a laser beam. The concept of this energy leveling can be applied to a coasting beam by adding one energy modulation at the beginning. The same principle can also be used to lower the beam energy spread and may find application in electron cooling~\cite{Budker1967Electron}. In this paper, we focus on using BBHG for density bunching. But we recognize that this energy distribution shaping may find important applications in the future.

\subsection{Weak-Strong SSMB Mechanism}

Having the appropriate microbunching method, now we need to solve the second key issue, i.e., making this high-harmonic bunching repeat turn-by-turn in a storage ring. A LWF SSMB storage ring can be invoked to provide a weakly bunched microbunch train as the first step. Then we do BBHG based on this microbunch train to generate extremely high-harmonic density bunching, which can be used for short-wavelength coherent radiation generation. Since the microbunch has been stretched to a length much longer than the laser wavelength at the modulator in BBHG, and considering the sinusoidal modulation waveform is nonlinear, we need to cancel this nonlinear modulation and de-stretch the beam to the initial length such that the weak focusing regime still apply to the whole beam to ensure the formation of the microbunch train in the first place. So we apply a negative dispersion $-r_{56}$ and a reverse laser modulation following the radiator to cancel the impact imprinted by the upstream modulation. After the modulation cancellation, we apply another negative dispersion $-R_{56}$ to remove the large chirp in longitudinal phase space distribution and recover the initial microbunch length.  This modulation and de-modulation process shares the spirit with the reversible seeding proposed in Ref.~\cite{Ratner2011Reversible}. But here the difference is that outside the reversible section, the beam is still microbunched which is essential to apply the BBHG manipulation. In this context, the proposed mechanism is a global SSMB scenario~\cite{Deng2025Stochastic}.


\begin{figure}[tb]
	\centering
	\includegraphics[width=0.85\linewidth]{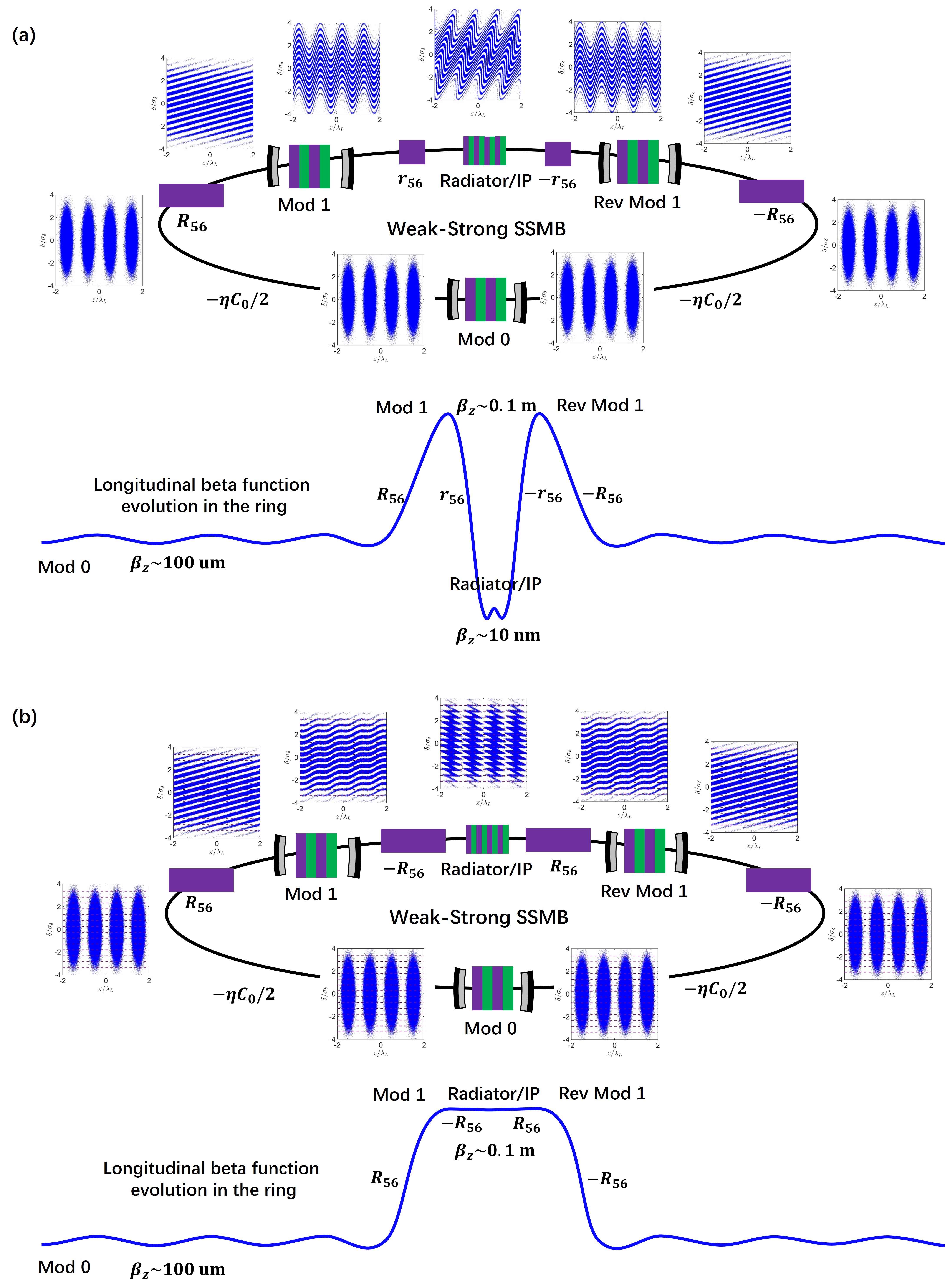}
	\caption{Schematic layouts of the proposed Weak-Strong SSMB storage ring. Also shown are the beam longitudinal phase space distribution and the sketch of longitudinal beta function evolution along the ring. (a) is for turn-by-turn high-harmonic density bunching, (b) is for turn-by-turn energy bunching.}
	\label{fig:weakstrongssmb}
\end{figure}

The schematic layout of and the beam longitudinal phase space distribution evolution in the proposed ring is shown in Fig.~\ref{fig:weakstrongssmb}a, where Mod represents laser modulator, and Rev Mod means reverse modulation.  We call this mechanism Weak-Strong SSMB. Although we have introduced the Weak-Strong mechanism primarily from a microbunching-centric viewpoint, its essence can also be appreciated from a storage ring beam dynamics perspective. To this end, we invoke the longitudinal Courant-Snyder formalism~\cite{Deng2021Courant}, in which the longitudinal beta function plays a key role (see Methods). From this perspective, the longitudinal optics control of this mechanism can be viewed as the longitudinal counterpart of a high-luminosity collider design, where the manipulation instead occurs in the transverse plane. The longitudinal beta function can vary by seven orders of magnitude along the ring, with its minimum compressed to the 10 nm level at around the radiator (or interaction point, IP). In fact, when operated as a light source, such a Weak-Strong SSMB ring can be interpreted as an electron-photon collider. Following the collider community, we refer to this strongly defocused and focused region as the final focus (FF).  Outside the FF region, $\beta_{z}$ is at $100\ \mu$m level as a result of the weak focusing of Mod 0 and the small storage ring momentum compaction. By applying a relatively large $R_{56}$, which can be viewed as a large longitudinal drift space, $\beta_{z}$ is boosted by three orders of magnitude to 0.1 m level at Mod 1. Mod 1 then acts as a focusing longitudinal quadrupole. And after another small longitudinal drift $r_{56}$, $\beta_{z}$ drops by seven orders of magnitude and is squeezed to the 10 nm level. The negative drifts $-r_{56}$ and $-R_{56}$, together with the defocusing kick from Rev Mod 1, then restore $\beta_{z}$ to the original 100~$\mu$m level. A sketch of the $\beta_{z}$ evolution along the ring is also presented in Fig.~\ref{fig:weakstrongssmb}.  Note that the above Courant-Snyder analysis applies when we linearize the laser modulation. The energy spread at the radiator for example is not dramatically increased considering the nonlinearity and finite amplitude of the sinusoidal modulation.



The key difference of this Weak-Strong scenario compared to the LSF SSMB~\cite{Chao2016SSMB,Zhang2021Ultra,Deng2021Courant,Deng2026TLCSSMB} is that now we have applied a large $|\pm R_{56}|$ to boost the $\beta_{z}$ by three orders of magnitude at the Mod 1 and Rev Mod 1. And the required focusing and defocusing strength and thus the modulation laser power is inversely proportional to $\beta_{z}$ (see Methods), and can be significantly lower. This is possible because we have adopted a modulation cancellation strategy in the final focus region, otherwise the bunch length at the modulator should be adequately smaller than the modulation laser wavelength to avoid strongly chaotic nonlinear dynamics~\cite{Chirikov1979}. This final focus region can be viewed as an extreme case of a LSF cell by pushing its synchrotron tune to zero, and handle over the global longitudinal focusing function in the ring to the weak focusing unit Mod 0. So the whole setup can be viewed as a special longitudinal strong focusing section embedded in a LWF SSMB ring, and this is why we call it Weak-Strong SSMB. 


Similarly the same mechanism can be used to create an energy-leveled beam turn-by-turn in a storage ring as shown in Fig.~\ref{fig:weakstrongssmb}b. More interesting applications of such a ring can be envisioned and we do not go into it in this paper.

\subsection{High-power EUV Source Based on Weak-Strong SSMB}\label{sec4}

As an example application of the Weak-Strong SSMB mechanism, we present a solution for a kW-level EUV source based on this scheme. The schematic layout of such a ring is shown in Fig.~\ref{fig:weakstrongssmb}a, and an example lattice layout is given in Fig.~\ref{fig:cell}. Key parameters of the storage ring are summarized in Tab.~\ref{tab1}. A few clarifying comments are in order. The beam energy is 600 MeV, which is adequate for EUV generation. A low energy makes intrabeam scattering (IBS) challenging, whereas an excessively high energy makes the laser modulation harder. The average beam current is 500 mA, and the peak current is 50~A, implying that 1\% of the ring circumference is filled with microbunches. A high peak current is desirable for achieving high average radiation power, given the quadratic dependence of the coherent radiation power on the beam current. The modulation laser wavelengths of the three modulators are all assumed to be 1030 nm, which lies in the mature wavelength range for high-power optical enhancement cavities (OECs). The average stored laser power in Mod 1 and Rev Mod 1 is assumed to be 1 MW, which is close to the state-of-the-art OEC technology~\cite{Lu2025OEC}. The laser power required for Mod 0 can be lower than this value. Note that this 1 MW is the stored power in the cavity; the injected laser power is at the 100 W level. The momentum compaction of the ring is $|\eta C_{0}|=100\ \mu$m, which is a small but achievable value. To speed up radiation damping and thus control the equilibrium beam emittances, damping wigglers with a peak field of 4 T and a total length of 52 m are employed. The method for controlling the quantum excitation of the transverse and longitudinal emittances in the damping wigglers is described in the Methods section and references therein. The total radiation power of the wigglers is about 100 kW. The energy compensation system of such a storage ring can be a conventional RF cavity, with harmonic cavities optionally added to tailor the beam current distribution. The microbunch length outside the final focus region is about 250 nm. The radiation wavelength is chosen as the $76$-th harmonic of the modulation laser, corresponding to about 13.5 nm. With a small amount of third-harmonic modulation applied to boost bunching (see Methods), the bunching factor at the radiator center is about 0.1. The radiator is assumed to be a planar undulator. With these parameters, the average EUV radiation power is 1 kW.

\begin{table}[tb]
	\caption{Example parameters of a Weak-Strong SSMB storage ring for EUV generation.}\label{tab1}%
	\begin{tabular}{@{}lll@{}}
		\toprule
		Parameter & Value  & Description \\
		\midrule
		$E_{0}$ & 600 MeV & Beam energy \\	
		$I_{\text{A}} $ & 500 mA & Average beam current \\
		$|\eta C_{0}|$ & 100 $\mu$m & Ring momentum compaction \\
		$|r_{56}|$ &  $50.1\ \mu\text{m}$ & Small $r_{56}$ \\ 
		$|R_{56}|$ &  $3.8$ mm & Large $R_{56}$ \\
		$B_{0{\text{W}}}$ &  4 T & Wiggler field strength\\
		$L_{{\text{W}}}$ &  52 m & Wiggler total length\\
		$P_{\text{W}}$ &  97 kW & Wiggler radiation power \\
		$\epsilon_{\bot}$ & 1 nm & Transverse emittance \\ 
		$\epsilon_{z}$ & 0.34 nm & Longitudinal emittance \\ 
		$\sigma_{\delta\text{B}}$ &  $1.4\times10^{-3}$ & Energy spread\\
		$\sigma_{z\text{B}}$ &  248 nm & Bunch length\\
		$\lambda_{\text{L}}$ & 1030 nm & Modulation laser wavelength	\\
		$P_{\text{L}}$ &  $\sim 1~\text{MW}$ & Average power\\ 
		$\lambda_{\text{R}}$ & $13.5$ nm & Radiation wavelength\\  
		$b_{76}$ & 0.095 & Bunching factor\\  
		$B_{0\text{R}}$ &  0.8 T &  Radiator field strength\\   
		$L_{u\text{R}}$ &  1.88 m  & Radiator length	\\  
		$P_{\text{R}}$ & 1 kW & Average EUV power \\                                          
		\botrule
	\end{tabular}
\end{table}

\section{Discussion}\label{sec13}

In summary, we have proposed a novel method for harmonic generation and energy bunching, termed BBHG, as well as a storage ring working mechanism, named Weak-Strong SSMB, which can efficiently manipulate the longitudinal beta function and thereby compress the bunch length. As an example application, we have presented a solution for a kW-level coherent EUV source based on the Weak-Strong SSMB mechanism. Our work fills a critical gap toward a storage ring-based, high-power, fully coherent short-wavelength light source. We anticipate that the proposed concepts may find broader applications beyond coherent radiation generation.

Some remarks on the critical issues and potential challenges of such a storage ring may be valuable for its further development. Although we aim to cancel the laser modulation of Mod 1 and Rev Mod 1, perfect cancellation is not achievable in practice, due to, for example, lattice nonlinearities, IBS, and coherent undulator radiation-induced distortion of the beam phase space distribution. This non-perfect cancellation has a direct impact on the required damping wiggler power. Given the large $\beta_{z}$ at these two modulators and the nonlinear nature of the modulation, the optimization of the nonlinear dynamics in such a ring--especially in the final focus region--requires dedicated effort. The collective beam dynamics in such a storage ring also require further in-depth study, given the breakdown of the adiabatic approximation--meaning that the longitudinal phase space evolves significantly along the ring--and the impact of turn-by-turn laser modulation. This is necessary to justify the beam parameters adopted in our solution. Some preliminary work in this direction has been carried out~\cite{Bian2025Threshold,Zhao2025Method,Pan2025IBS}, and the results are encouraging. Another critical issue is phase locking among the laser modulators--or, more precisely, between the laser and the electron arrival time. Further work is ongoing and will be reported in the future.

\clearpage

\section*{Methods}\label{sec11}

\subsection*{BBHG Density Bunching Factor}
\subsubsection*{Singe bunch}


In this paper, we use $(z,\delta)^{T}$ as the particle state vector in longitudinal phase space, where $z$ means the longitudinal position and $\delta=(E-E_{0})/E_{0}$ means the relative energy deviation with respect to the reference particle, $E_{0}$ is the reference particle energy. A laser modulator imprints an energy modulation $\delta=\delta+A_{\text{L}}\sin(k_{\text{L}}z)$, and a dispersion section shifts the particle longitudinal coordinate according to its energy $z=z+R_{56}\delta$. The density bunching factor of BBHG as shown in Fig.~\ref{fig:BBHG}a can be derived
\begin{equation}\label{eq:BFSingle}
\begin{aligned}
b(k_{z})&=\langle e^{-ik_{z}z}\rangle
=\sum_{p=-\infty}^{\infty}J_{p}(-k_{z}r_{56}A_{\text{L}})\text{exp}\left[-\frac{\epsilon_{z}}{2}\left(\beta_{z}M^{2}-2\alpha_{z}MN+\gamma_{z}N^{2}\right)\right],\\
\end{aligned}
\end{equation}
with $M\equiv (k_{z}-pk_{\text{L}})$, $N\equiv (k_{z}-pk_{\text{L}})R_{56}+k_{z}r_{56}$, $p$ being an integer, $R_{56}$ and $r_{56}$ are the first- and second-stage dispersion section strength, respectively. We have assumed that the initial bunch distribution is Gaussian, and $\alpha_{z}$, $\beta_{z}$, $\gamma_{z}$ are the initial longitudinal Courant-Snyder parameters~\cite{Deng2021Courant} before entering the $R_{56}$ dispersion section.
If $k_{\text{L}}\sigma_{z\text{B}}\gg\pi$, then there is bunching  only when $k_{z}=nk_{\text{L}}$ with $n$ being an integer, and there is only one term  $p=n$ contributes significantly in the summation, 
\begin{equation}
\begin{aligned}
b_{n}=b(k_{z}=nk_{\text{L}})
&\approx J_{n}(-nk_{\text{L}}r_{56}A_{\text{L}})\text{exp}\left[-{\left(nk_{\text{L}}r_{56}\sigma_{\delta \text{B}}\right)^{2}}/{2}\right],
\end{aligned}
\end{equation}
which as expected is the classical bunching factor of HGHG~\cite{Yu1991} based on a coasting beam. $\sigma_{z\text{B}}=\sqrt{\epsilon_{z}\beta_{z}}$ and $\sigma_{\delta\text{B}}=\sqrt{\epsilon_{z}\gamma_{z}}$ are the initial bunch length and energy spread, respectively, with $\epsilon_{z}$ the longitudinal emittance of the beam.

Now we consider our interested case $k_{\text{L}}\sigma_{z\text{B}}\lesssim\pi$.
For simplicity and without loss of generality, we assume $\alpha_{z}=0$. If $|k_{\text{L}}R_{56}\sigma_{\delta\text{B}}|\gg1$, and denote $R\equiv\frac{R_{56}}{r_{56}}$, $m\equiv\frac{R+1}{n+1}$,  then for $k_{z}=\frac{R}{m}k_{\text{L}}$, only one term of $p=n+1$ contributes significantly,
%
%
\begin{equation}\label{eq:BFnSingle}
b\left(k_{z}=\frac{R}{m}k_{\text{L}}\right)\approx J_{n+1}\left(-k_{z}r_{56}A_{\text{L}}\right)\text{exp}\left[-{\left(\frac{k_{\text{L}}\sigma_{z\text{B}}}{m}\right)^{2}}/{2}\right].
\end{equation}
For example if $R=n$, which means  $m=1$, then we have
\begin{equation}
b_{n}\approx J_{n+1}(-nk_{\text{L}}r_{56}A_{\text{L}})\text{exp}\left[-{\left(k_{\text{L}}\sigma_{z\text{B}}\right)^{2}}/{2}\right].
\end{equation}
For $n>4$, let
$
|nk_{\text{L}}r_{56}A_{\text{L}}|=n+1+0.81(n+1)^{1/3},
$ we have the maximal bunching, 
\begin{equation}
\begin{aligned}
b_{n}
&\approx\frac{0.67}{(n+1)^{1/3}}\text{exp}\left[-{(k_{\text{L}}\sigma_{z\text{B}})^2}/{2}\right].
\end{aligned}
\end{equation}
Note that this bunching factor is independent of the initial energy spread $\sigma_{\delta\text{B}}$ and the energy modulation $A_{\text{L}}$. Also note that the exponential term does not depend on the harmonic number $n$. If we can make $k_{\text{L}}\sigma_{z\text{B}}\lesssim1$, then the bunching factor can approach the theoretical maximal ${0.67}/{(n+1)^{1/3}}$. Even if $k_{\text{L}}\sigma_{z\text{B}}$ is larger, we can implement a larger $|m|>1$ to boost the bunching. So in principle, BBHG applied to a single bunch can realize a theoretical maximal bunching factor, while the required energy modulation strength can be arbitrarily weak.

\begin{figure}[tb]
	\centering
	\includegraphics[width=1\linewidth]{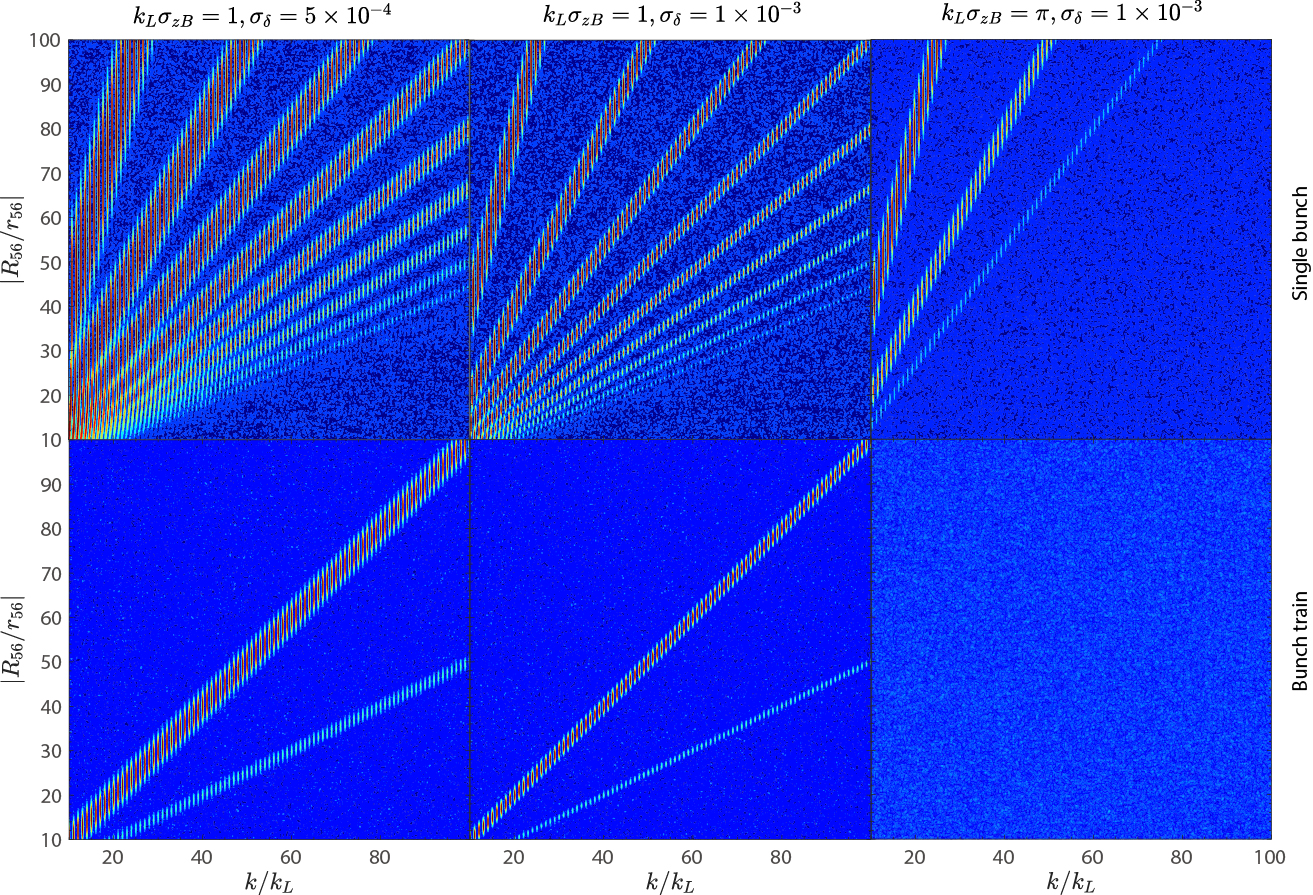}
	\caption{Contour plot of $\sqrt{|b(k)|}$ vs $(k/k_{\text{L}},|{R_{56}}/{r_{56}}|)$, with $r_{56}=\frac{n+1+0.81(n+1)^{1/3}}{nk_{\text{L}}A_{\text{L}}}$ where $n=76$. 
		We have used $\sqrt{|b(k)|}$, instead of $|b(k)|$, to increase the contrast of illustration in the plot. 
	}
	\label{fig:bf2dsinglelowes}
\end{figure}

\subsubsection*{Bunch train}
Now we consider the case of BBHG on a bunch train with a period of $\lambda_{\text{B}}$. In this paper, we use the subscript S and T to represent single bunch and bunch train, respectively. The bunching factor of train-based BBHG can be derived,
\begin{equation}
\begin{aligned}
b_{\text{T}}(k_{z})&=b_{\text{S}}(k_{z})\sum_{q=-\infty}^{\infty}\delta(k_{z}-pk_{\text{L}}-qk_{\text{B}}),\\
\end{aligned}
\end{equation}
where $b_{\text{S}}(k_{z})$ is given in Eq.~(\ref{eq:BFSingle}), $q$ is an integer, $k_{\text{B}}=2\pi/\lambda_{\text{B}}$ and
$
\delta(x)=\begin{cases}
1,\ x=0,\\
0,\ \text{else}.
\end{cases}
$
Note that if $\lambda_{\text{B}}/\lambda_{\text{L}}$ is not an integer, the current distribution of each bunch after the BBHG manipulation is not identical.

To get a non-vanishing bunching factor, we need $k_{z}=pk_{\text{L}}+qk_{\text{B}}$. 
Denote $\kappa\equiv \frac{k_{\text{L}}}{ k_{\text{B}}}$,
 we can choose $m=-\frac{\kappa}{q}$, $R=-\frac{\kappa}{q}(n+1)-1$, and then 
 \begin{equation}
 \begin{aligned}
 b_{\text{T}}\left(k_{z}=\left(n+1+\frac{q}{\kappa}\right)k_{\text{L}}\right)=b_{\text{S}}(k_{z})
 &\approx J_{n+1}\left(-k_{z}r_{56}A_{\text{L}}\right)\text{exp}\left[-{\left(qk_{\text{B}}\sigma_{z\text{B}}\right)^{2}}/{2}\right].
 \end{aligned}
 \end{equation}
For example, if $\kappa=1$, $q=-1$, $R=n$, then $b_{n}$ is given by Eq.~(\ref{eq:BFnSingle}).  In finding the parameters given in Tab.~\ref{tab1}, we have assumed that Mod 1 uses the same laser wavelength as that of Mod 0, therefore $\kappa=1$. Note that we can also apply a $\kappa\neq1$.
From the above derivation we know that it is $k_{\text{B}}\sigma_{z\text{B}}$, i.e., the initial bunching degree at the train period,  that determines the upper limit of the final bunching factor at high harmonics, which is reasonable. Note also that for the case of bunch train-based BBHG, we cannot apply a large $|m|>\kappa$ to weaken the impact of $k_{\text{B}}\sigma_{z\text{B}}$ in the exponential term as that in the case of a single bunch. This is because $|q|\geq1$. All the derivations and observations have been confirmed by the numerical calculations of bunching factor as shown in Fig.~\ref{fig:bf2dsinglelowes}.

\subsection*{BBHG Energy Bunching Factor}
As shown in Fig.~\ref{fig:BBHG}b, BBHG can also used for energy bunching, creating energy levels in a bunched beam. 
The energy difference of the neighboring levels is given by
$ 
\Delta\delta={\lambda_{\text{L}}}/{R_{56}},
$
and can be flexibly adjusted by the laser wavelength $\lambda_{\text{L}}$ and the dispersion section strength $R_{56}$.
Similar to the density bunching, the BBHG energy bunching factor of a single bunch can be derived
\begin{equation}
\begin{aligned}
\mathcal{B}_{\text{S}}(k_{\delta})&=\langle e^{-ik_{\delta}\delta}\rangle
=\sum_{p=-\infty}^{\infty}J_{p}(-k_{\delta}A_{\text{L}})\text{exp}\left[-\frac{\epsilon_{z}}{2}\left(\beta_{z}P^{2}-2\alpha_{z}PQ+\gamma_{z}Q^{2}\right)\right],
\end{aligned}
\end{equation}
with $P\equiv pk_{\text{L}}$, $Q\equiv pk_{\text{L}}R_{56}-k_{\delta}$. In this paper, we use $b$ and $\mathcal{B}$ to distinguish the density and energy bunching factor. Note that the energy bunching factor is independent of the second-stage dispersion $r_{56}$. Denote 
$
k_{\Delta\delta}\equiv k_{\text{L}}R_{56}
$, and assume that $\sigma_{\delta\text{B}}k_{\Delta\delta}\gg1$. For $k_{\delta}=nk_{\Delta\delta}$,  then only one term of $p=n$ contributes significantly in the above summation, and we have
\begin{equation}
\begin{aligned}
\mathcal{B}_{\text{S}}(k_{\delta}=nk_{\Delta\delta})
&\approx J_{n}(-nk_{\Delta\delta}A_{\text{L}})\text{exp}\left[-{(nk_{\text{L}}\sigma_{z\text{B}})^{2}}/{2}\right].
\end{aligned}
\end{equation}
To maximize the energy bunching with $n=1$, we can make $|k_{\Delta\delta}A_{\text{L}}|=1.84$.
Similarly, the energy bunching factor for a bunch train-based BBHG is
\begin{equation}
\begin{aligned}
\mathcal{B}_{\text{T}}(k_{\delta})&=\mathcal{B}_{\text{S}}(k_{\delta})\sum_{q=-\infty}^{\infty}\delta(pk_{\text{L}}-qk_{\text{B}}).
\end{aligned}
\end{equation}

\subsection*{Enhance Bunching by Addition of High-Harmonic Modulation}

We can add high-harmonic modulations in addition to the fundamental-frequency one to boost the final bunching factor. For example, if we add a third-harmonic modulation in addition to the fundamental frequency with zero phase shift with respect to each other, we have the density bunching factor
\begin{equation}
\begin{aligned}
b_{n}
&=\sum_{p_{1}=-\infty}^{\infty}J_{p_{1}}( \theta A_{1})\sum_{p_{3}=-\infty}^{\infty}J_{p_{3}}(\theta A_{3})\text{exp}\left[-\frac{\epsilon_{z}}{2}\left(\beta_{z}V^{2}-2\alpha_{z}VY+\gamma_{z}Y^{2}\right)\right],
\end{aligned}
\end{equation}
with $\theta\equiv-nk_{\text{L}}r_{56}$, $A_{1}$ and $A_{3}$ being the modulation strength of the fundamental frequency and the third harmonic, receptively,  $V\equiv (n-p_{1}-3p_{3})k_{\text{L}}$, $Y\equiv [(n-p_{1}-3p_{3})R_{56}+nr_{56}]k_{\text{L}}$. 
For $n=76$, the optimal  ${A_{3}}/{A_{1}}\approx-0.064$ and the 76-th laser harmonic can be a factor 1.92 larger than the case without the third-harmonic laser. In Tab.~\ref{tab1}, we have used this bunching boost. The same method can also be used to enhance the energy bunching. A planar undulator can be used as the modulator simultaneously for the fundamental and third-harmonic modulation. Note that this bunching factor enhancement is not mandatory in the proposed Weak-Strong SSMB mechanism.





\subsection*{Longitudinal Courant-Snyder Analysis}

\subsubsection*{Weak-Strong SSMB}


In a planar uncoupled ring, the equilibrium longitudinal emittance given by the balance of quantum excitation and radiation damping is~\cite{Chao1979SLIM}
\begin{equation}
\epsilon_{z}=\frac{C_{L}\gamma^{5}}{2c\alpha_{L}}\oint\frac{\beta_{z}}{|\rho|^{3}}ds
\end{equation}
where $C_{L}=\frac{55}{48\sqrt{3}}\frac{r_{e}\hbar}{m_{e}}$ with $r_{e}$ the classical electron radius, $\hbar$ the reduced Planck's constant, $m_{e}$ the electron mass, $\gamma$ is the Lorentz factor, $c$ is the speed of light in free space,  $\rho$ is the bending radius, and $\beta_{z}$ is the longitudinal beta function, $\alpha_{L}=\frac{U_{0}}{2E_{0}}J_{s}$ is the longitudinal damping constant,  $U_{0}$ is the radiation loss per particle per turn, and $J_{s}$ is the longitudinal damping partition number, and nominally $J_{s}\approx2$. 

The longitudinal beta function can be obtained by doing Courant-Snyder parametrization~\cite{Courant1958Theory} in the longitudinal dimension~\cite{Deng2021Courant}. The RF or laser modulation after linearization is like a longitudinal quadrupole whose transfer matrix is $\left(\begin{matrix}
1&0\\
h&1
\end{matrix}\right)$ with $h=\frac{eV_{\text{L}}\sin\phi}{E_{0}}k_{\text{L}}$, while the dispersion section is like a longitudinal drift space whose transfer matrix is $\left(\begin{matrix}
1&R_{56}\\
0&1
\end{matrix}\right)$. If the modulator momentum compaction is also taken into account, the transfer matrix of a laser modulator then corresponds to a thick-lens longitudinal quadrupole~\cite{Deng2024Springer}. The magnetic lattice outside the final focus region of the Weak-Strong SSMB (see Figs.~\ref{fig:weakstrongssmb} and \ref{fig:cell}) has a dispersion strength $-\eta C_{0}$, where $\eta$ and $C_{0}$ are the storage ring phase slippage factor and circumference, respectively. The one-turn map ${\bf M}$ is obtained by multiplying the transfer matrices of each section of the ring in order. Then we can do the parameterization on ${\bf M}$
\begin{equation}
{\bf M}=\left(\begin{matrix}
\cos\Phi_{s}+\alpha_{z}\sin\Phi_{s}&\beta_{z}\sin\Phi_{s}\\
-\gamma_{z}\sin\Phi_{s}&\cos\Phi_{s}-\alpha_{z}\sin\Phi_{s}
\end{matrix}\right),
\end{equation}
where $\Phi_{s}=2\pi\nu_{s}$ is the global synchrotron phase advance per turn, and $\nu_{s}$ is the synchrotron tune.
When $0<h_{0}\eta C_{0}\ll1$, which means the global synchrotron turn is much less than 1, the longitudinal beta function at the Mod 0 in the proposed Weak-Strong SSMB is
\begin{equation}
\begin{aligned}
\beta_{z\text{M0}}\approx\sqrt{\frac{\eta C_{0}}{h_{0}}}\approx \frac{|\eta C_{0}|}{2\pi\nu_{s}},
\end{aligned}
\end{equation}
where $h_{0}$ is the linear energy chirp imprinted by Mod 0, $\nu_{s}\approx \frac{\sqrt{h_{0}\eta C_{0}}}{2\pi}$ is the global synchrotron tune. In finding the parameters given in Tab.~\ref{tab1}, we have ensured that $h_{0}\eta C_{0}\leq0.3$ to avoid strong chaos in the phase space dynamics~\cite{Chirikov1979}.
The longitudinal beta function right before $R_{56}$ is
$
\beta_{z0}=\left(1-{h_{0}\eta C_{0}}/{4}\right)\beta_{z\text{M0}}\approx\beta_{z\text{M0}}.
$
For simplicity, we have assumed that Mod 0 is at a symmetric location with respect to Mod 1 and Rev Mod 1, which is not compulsory from a beam dynamics viewpoint.
The longitudinal beta function at Mod 1 and Rev Mod 1 is
\begin{equation}
\begin{aligned}
\beta_{z\text{M1}}= \beta_{z0}+\frac{R_{56}^{2}}{\beta_{z0}}.
\end{aligned}
\end{equation}
After the Mod 1 kick and the transport of the dispersion section $r_{56}$, the longitudinal beta function at the radiator or interaction point is
 $
 \beta_{z\text{R}}= 
 \beta_{z0}\zeta^{2}
 + \frac{r_{56}^{2}\left(1+ R\zeta \right)^{2}}{\beta_{z0}},
 $
 where we have used the denotation $\zeta\equiv 1+h_{1}r_{56}$, $R\equiv R_{56}/r_{56}$.
 When
 $
 \zeta=-1/[R+\frac{1}{R}\left({\beta_{z0}}/{r_{56}}\right)^{2}],
 $
 we have the minimal $\beta_{z\text{R}}$,
 \begin{equation}
 \beta_{z\text{R},\text{min}}
 =\frac{\beta_{z0}}{R^{2}+\left({\beta_{z0}}/{r_{56}}\right)^{2}}.
 \end{equation}
 So $\beta_{z\text{R}}$ can be efficiently squeezed by  applying a large $|R\equiv R_{56}/r_{56}|$, and the required energy chirp strength $|h_{1}|$ and thus the modulation laser power can be flexibly lowered by increasing the magnitudes of $R_{56}$ and $r_{56}$. This is the essence of the proposed Weak-Strong mechanism.

 With the squeezed $\beta_{z\text{R}}$, the linear bunch length $\sigma_{z\text{R}}=\sqrt{\epsilon_{z}\beta_{z\text{R}}}$ at the radiator, which means if we consider only the linearized part of the laser modulation, correspondingly can be more than a factor of $|R|$ smaller than that of the initial bunch $\sigma_{z\text{B}}=\sqrt{\epsilon_{z}\beta_{z0}}$, thus allowing large bunching factor at short wavelengths. When $|R|\gg1$ and with the optimal bunch compression, we have
 $
 h_{1}^{2}\beta_{z\text{M1}}\beta_{z\text{R}}\approx 1.
 $
 The required laser power for Mod 1, 
 \begin{equation}
 P_{\text{LM1}}\propto h_{1}^{2}\approx \frac{1}{\beta_{z\text{M1}}\beta_{z\text{R}}}=\left(\frac{\sigma_{\delta\text{M1},\text{s}}}{\sigma_{z\text{R}}}\right)^{2},
 \end{equation}
 where $\sigma_{\delta\text{M1},\text{s}}=\sqrt{\epsilon_{z}/\beta_{z\text{M1}}}$ is the slice energy spread at Mod 1, $\sigma_{z\text{R}}$ is determined by our desired coherent radiation wavelength. So to lower the required modulation laser power with a given desired coherent radiation wavelength, we need to lower the slice energy spread at the Mod 1 by enlarging $\beta_{z}$ there, which is exactly what the large $R_{56}$ is used for. By implementing a large $R_{56}$, we enlarge $\beta_{z}$ by three orders of magnitude at Mod 1 compared to that before $R_{56}$, which in turn results a three orders of magnitude drop in the required laser power.

\begin{figure}[tb]
	\centering
	\includegraphics[width=0.6\linewidth]{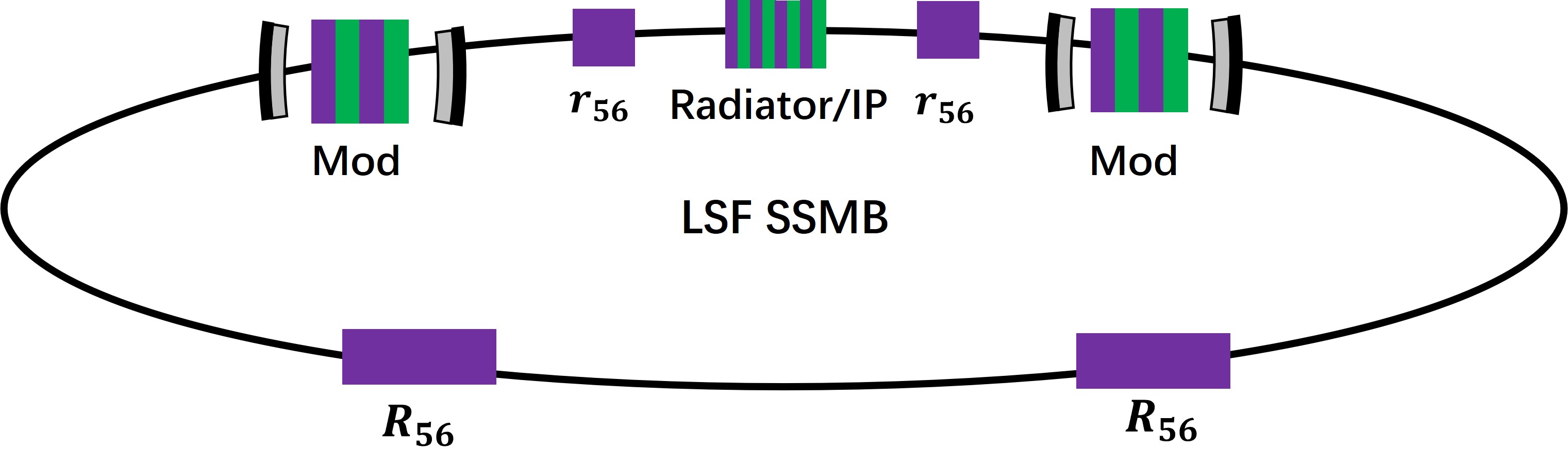}
	\caption{An example schematic layout of a LSF SSMB storage ring.}
	\label{fig:lsfssmb}
\end{figure}

\subsubsection*{Comparison with LSF SSMB}


To better appreciate the essence and advantage of the proposed Weak-Strong SSMB, for comparison here we also present a short analysis of the LSF SSMB whose schematic layout is shown in Fig.~\ref{fig:lsfssmb}. Denote $\zeta_{1}\equiv 1+hr_{56}$, $\zeta_{2}\equiv 1+hR_{56}$, the one-turn map at the radiator, modulator and radiator opposite are respectively given by
 \begin{equation}
 \begin{aligned}
 {\bf M}_{\text{R}}&=\left(
 \begin{array}{cc}
 2 \zeta _1 \zeta _2-1 & \frac{2 \zeta _1 \left(\zeta _1 \zeta _2-1\right)}{h} \\
 2 \zeta _2 h & 2 \zeta _1 \zeta _2-1 \\
 \end{array}
 \right),\
 {\bf M}_{\text{M}}=\left(
 \begin{array}{cc}
 2 \zeta _2-1 & \frac{4\zeta_{1}\zeta_{2}-2(\zeta_{1}+\zeta_{2})}{h} \\
 2 \zeta _2 h & \left(4 \zeta _1-2\right) \zeta _2-1 \\
 \end{array}
 \right),\\
 {\bf M}_{\text{RO}}&=\left(
 \begin{array}{cc}
 2 \zeta _1 \zeta _2-1 & \frac{2 \zeta _2 \left(\zeta _1 \zeta _2-1\right)}{h} \\
 2 \zeta _2 h & 2 \zeta _1 \zeta _2-1 \\
 \end{array}
 \right).\\
 \end{aligned} 
 \end{equation} 
 Linear stability requires $0<\zeta_{1}\zeta_{2}<1$. Like that in Weak-Strong SSMB, we require that $\beta_{z\text{M}}$ can be three orders of magnitude larger than $\beta_{z\text{RO}}$ which in turn can be three orders of magnitude larger than $\beta_{z\text{R}}$. The first requirement is to lower the required modulation laser power, and the second requirement is to realize an ultrashort bunch length at the radiator. We assume that $\beta_{z}$ in the main part of the ring, which is the section between two $R_{56}$ in Fig.~\ref{fig:lsfssmb}, has a similar level with $\beta_{z\text{RO}}$, such that to control the equilibrium longitudinal emittance. To realize these requirements, we need  $|\zeta_{1}|\approx \frac{1}{\sqrt{1000}}$, $|\zeta_{2}|\approx \sqrt{1000}$, $1-\zeta_{1}\zeta_{2}=\frac{1}{1000}$, $\nu_{s}\approx1-\frac{1}{\pi\sqrt{1000}}$. Then 
 $
 \beta_{z\text{R}}=\frac{1}{1000h},\
 \beta_{z\text{M}}=\frac{1000}{h},\
 \beta_{z\text{RO}}=\frac{1}{h}.
 $
 We remind that in this case we also have $
 h^{2}\beta_{z\text{M}}\beta_{z\text{R}}\approx 1
 $. Actually this relation  is universal in strong focusing-related bunch compression or harmonic generation schemes~\cite{Deng2026TLCSSMB}. For GLSF or transverse-longitudinal coupling based SSMB, this relation is $h^{2}\mathcal{H}_{y\text{M}}\mathcal{H}_{y\text{R}}\geq1$, where $\mathcal{H}$ is the chromatic function which quantifies the impact of transverse emittance on longitudinal bunch length. If $h=10^{4}\ \text{m}^{-1}$, then the $\beta_{z}$ at the three locations in such a LSF SSMB ring can reach similar values as that in the proposed Weak-Strong SSMB.
 So from a linear dynamics perspective, LSF SSMB in principle can also realize the large $\beta_{z}$ ratios between these three locations. But the problem is that when we consider the fact that the sinusoidal modulation is nonlinear, it is not feasible to apply such a large $\beta_{z\text{M}}$ at the modulator in LSF SSMB since then it breaks the condition $\sigma_{z}=\sqrt{\epsilon_{z}\beta_{z}}< \lambda_{\text{L}}$ and nonlinear dynamics will not permit~\cite{Chirikov1979}. While in the proposed Weak-Strong SSMB, since the nonlinear modulation waveform of Mod 1 and Rev Mod 1 has been canceled, it allows us to break this condition at the two modulators. We can apply a much larger $\beta_{z}$ there, and thus lower the required modulation laser power notably. The final focus region in the Weak-Strong SSMB in this sense can be viewed as an extreme case of a LSF section by pushing the synchrotron tune to zero. With this modulation cancellation, we need a separate focusing unit to bunch the beam in the first place, which is what Mod 0 is used for. So the whole setup can be viewed as a special longitudinal strong focusing section embedded in a LWF SSMB ring, and this is why we call it Weak-Strong SSMB.

 

\subsubsection*{Control of longitudinal beta function}

\begin{figure}[tb]
	\centering
	\includegraphics[width=1\linewidth]{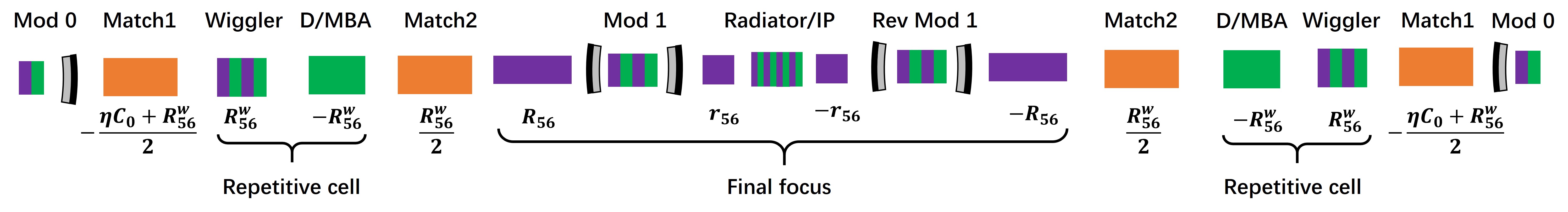}
	\caption{An example lattice layout to control $\beta_{z}$ evolution in the proposed Weak-Strong SSMB ring. 
	}
	\label{fig:cell}
\end{figure}

According to the bunching factor formula, we know that to get a large bunching at high harmonics, we need $k_{\text{L}}\sigma_{z\text{B}}\lesssim\pi/2$.  And from the above analysis we know that the key in realizing this is controlling the longitudinal beta function in the ring to realize a small longitudinal emittance and bunch length. An example layout of the lattice design for such a Weak-Strong SSMB ring is presented in Fig.~\ref{fig:cell}. In this design, we have made the repetitive cell consisting of a wiggler and a double or multi bend achromat (D/MBA) isochronous to minimize the variation of partial or local momentum compaction, thus $\beta_{z}$, outside the final focus region. The two matching sections are also for this purpose, in addition to ensure the required global phase slippage of the ring. There could be other designs as long as we can realize the required small longitudinal emittance and bunch length. Some more indepth analysis on minimization of longitudinal emittance can be found in Refs.~\cite{Deng2021Courant,Zhang2021Ultra,Deng2026TLCSSMB}.

As will be shown in the next section, for high-power short-wavelength coherent radiation generation, we may need to apply damping wigglers in the ring to speed up radiation damping.  The damping wiggler however also contributes to quantum excitation. To control the  transverse emittance~\cite{Deng2026TLCSSMB},  we need to tailor the transverse optics in the wigglers and the wiggler period length 
$
\lambda_{\text{W}}[\text{m}]\leq 3.19\sqrt{\frac{N_{\text{W}c}J_{x}E_{0}[\text{GeV}]\epsilon_{x0}[\text{nm}]}{B_{0\text{W}}^{3}[\text{T}]L_{\text{W}}[\text{m}]}},
$
where we have assumed that the total length of the damping wigglers are the sum of $N_{wc}$ identical ones and the lattice optics in them are also identical. 
Nominally the horizontal damping partition number $J_{x}\approx1$. For our example parameters, $E_{0}=600$~MeV,  $\epsilon_{x0}=1$~nm,  $B_{0\text{W}}=4$ T, $L_{\text{W}}=52$~m, $N_{\text{W}c}=26$ which means each small wiggler has a length of 2~m, then we need
$
\lambda_{\text{W}}\leq22\ \text{cm},
$ which is not a stringent requirement. 

To minimize the wiggler momentum compaction thus to minimize $\beta_{z}$ variation in it, we will use as small $\lambda_{\text{W}}$ as allowed. A wiggler with a period length of $5$~cm and peak field of 4 T is doable using superconducting magnet technology~\cite{Shkaruba2018Wiggler}. If $B_{\text{W}}=4$~T, $\lambda_{\text{W}}=5$ cm,  the resonance wavelength is
$
\lambda_{\text{W}R}=\frac{1+{K_{\text{W}}^{2}}/{2}}{2\gamma^{2}}\lambda_{\text{W}}=3.18\ \mu\text{m},
$
where $K_{\text{W}}=0.934\cdot B_{\text{W}}[\text{T}]\cdot\lambda_{\text{W}}[\text{cm}]$ is the undulator parameter determined by the undulator or wiggler field strength and period length.
For each wiggler we have a period number $N_{\text{W}}=40$, then the wiggler momentum compaction is
$
R_{56}^{\text{W}}=2N_{\text{W}}\lambda_{\text{W}R}=254.6\ \mu\text{m}.
$
Assume that we have compensated the $R_{56}^{\text{W}}$ symmetrically at the two sides of each wiggler as shown in Fig.~\ref{fig:cell}, then $\beta_{z}$ at the entrance and exit of each wiggler is
$
\beta_{z,\text{WE}}\approx\beta_{z0}+\frac{\left({{R_{56}^{\text{W}}}/{2}}\right)^{2}}{\beta_{z0}}=270\ \mu\text{m},
$ with $\beta_{z0}=181\ \mu$m.
$\beta_{z}$ outside the final focus region will oscillates between  $\beta_{z\text{M0}}$ and $\beta_{z,\text{WE}}$.  Since $\beta_{z}$ at Mod 1 and Rev Mod 1 are large, the quantum excitation from the two laser modulators should also be carefully evaluated. To realize a small longitudinal emittance, generally we prefer a weak bending magnetic field where $\beta_{z}$ is large. For a planar undulator modulator, the quantum excitation of energy spread is $\Delta\sigma_{\delta}^{2}=1.41\times10^{-12}E_{0}^{2}[\text{GeV}]B_{0}^{3}[\text{T}]L_{u}[\text{m}]$. We have applied a modulator field strength of about 1 T and a length of meter level, so the quantum excitation of energy spread square is three orders of magnitude weaker than that outside the final focus region.  We have assumed that $\langle \beta_{z}\rangle=1.2\beta_{z\text{M0}}$ in finding the parameters in Tab.~\ref{tab1} and this should be reasonable. Note that this averaging should be weighted at bending-related elements around the ring.

\subsection*{Requirement on Radiation Damping}

Note that in the proposed Weak-Strong SSMB mechanism, apart from quantum excitation, there is another source of longitudinal emittance growth, i.e.,  the non-perfect modulation cancellation between Mod 1 and Rev Mod 1. 
Assuming there is an effective rms longitudinal coordinate deviation $\sigma_{\Delta z}$ for the beam each time passing the section from Mod 1 to Rev Mod~1, then the energy spread growth after traversing the Rev Mod 1 is
$
\Delta \sigma_{\delta,\text{M1}}^{2}=\frac{(h_{1}\sigma_{\Delta z})^{2}}{2}.
$
The equilibrium longitudinal emittance considering this contribution is
\begin{equation}
\epsilon_{z}=\frac{\Delta\epsilon_{z,\text{ring}}+\Delta\epsilon_{z,\text{FF}}}{2\alpha_{L}},
\end{equation}
with $\Delta\epsilon_{z,\text{ring}}=\oint\frac{d\Delta\sigma_{\delta,\text{QE}}^{2}}{ds}\frac{\beta_{z}(s)}{2}ds=\Delta \sigma_{\delta,\text{QE}}^{2}\langle \beta_{z}\rangle/2$ and $\Delta\epsilon_{z,\text{FF}}=\Delta \sigma_{\delta,\text{M1}}^{2}\beta_{z\text{M1}}/2$ representing the excitation of longitudinal emittance from quantum excitation of the whole ring, and that of the final focus section from $\sigma_{\Delta z}$ per turn, respectively.

Denote $\Omega\equiv\frac{\Delta\epsilon_{z,\text{ring}}}{\Delta\epsilon_{z,\text{FF}}}$,  $\Gamma\equiv k_{\text{L}}\sigma_{z\text{B}}$, $\Lambda\equiv k_{\text{R}}\sigma_{\Delta z}$, then we have
\begin{equation}
\begin{aligned}
\Gamma&\approx \sqrt{\frac{(1+\Omega)N_{z}}{8}} \Lambda,\
\sigma_{\Delta z} \approx  \frac{1}{\sqrt{\pi (1+\Omega)/\Gamma }}\frac{\lambda_{\text{R}}}{\sqrt{N_{z}}},\
U_{0}\approx \frac{(1+\Omega) E_{0}}{4\pi\Gamma}\Lambda^{2},
\end{aligned}
\end{equation}
where $N_{z}=1/\alpha_{L}\approx E_{0}/U_{0}$ is the longitudinal radiation damping time in unit of revolution number. 
%
%
In our example parameters given in Tab.~\ref{tab1}, $E_{0}=600$ MeV, ${\sigma_{\Delta z}}=0.14$ nm, $\lambda_{\text{R}}=13.5$ nm, $\Lambda=0.064$, $\Omega=0.38$, $\Gamma=1.5$, so we have $U_{0}\approx 180$ keV, which is consistent with the damping wiggler power given in the table. 

Now let us consider the scaling of coherent radiation power on the damping strength with all the other parameters given,
$
P_{\text{R}}\propto |b|^{2}\propto \text{exp}(-{a}/{U_{0}}),
$
where we have used the denotation $a\equiv \frac{(1+\Omega)\Lambda^{2}E_{0}}{8}$. We recognize that $\Omega$ actually has an implicit dependence on $U_{0}$, and that the effective $\sigma_{\Delta z}$ also has an implicit dependence on the radiation power considering the coherent radiation induced phase space distortion.
Having considered these points, there is a specific $U_{0}$ where the ratio of coherent radiation energy and radiation damping loss reaches the maximal.

Some comments are in order concerning the effective $\sigma_{\Delta z}$, which has a direct impact on the damping wiggler power required. Some important physical effects leading to the effective $\sigma_{\Delta z}$ include lattice nonlinearity, intrabeam scattering (IBS), and coherent undulator radiation-induced energy spread growth. We recognize that dedicated lattice design efforts and IBS optimization are needed to realize the applied $\sigma_{\Delta z}=0.14$~nm. Some innovation lattice design for this purpose can be found in Ref.~\cite{Pan2025Isochronous}. To mitigate IBS, we can apply a transversely round beam. Concerning the coherent undulator radiation-induced distortion, we may implement a small negative momentum compaction between two radiators and let the radiation from the upstream radiator act back on the electron beam in the downstream one to minimize the energy spread growth from radiation process. If the $\sigma_{\Delta z}$ realizable is larger or smaller than that used in our example, the required damping wiggler power will correspondingly higher or lower than presented here.

\begin{figure}[tb]
	\centering
	\includegraphics[width=0.85\linewidth]{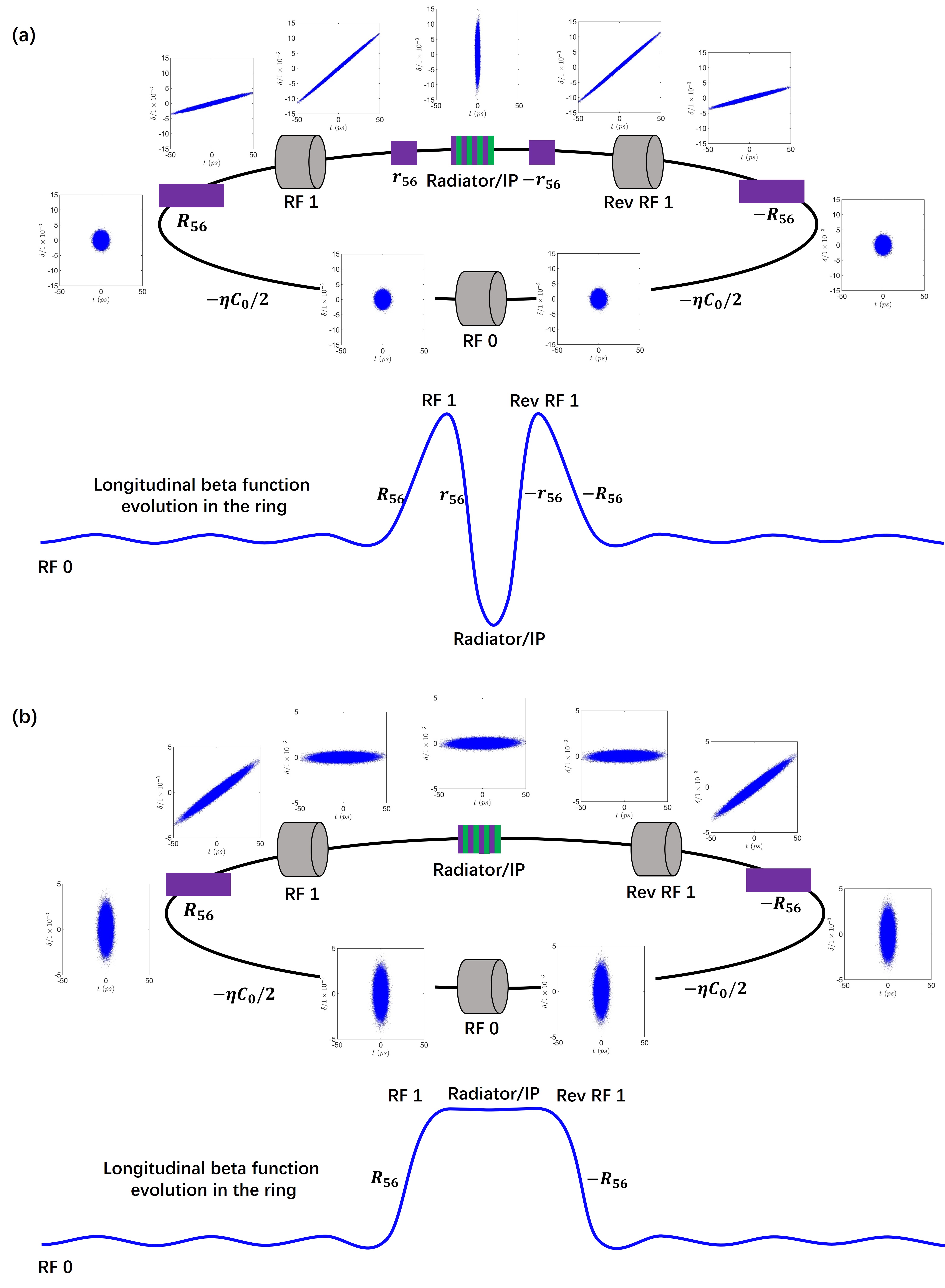}
	\caption{Schematic layout of the Weak-Strong mechanism at RF wavelength for bunch length (a) and energy spread (b) control in a storage ring.}
	\label{fig:weakstrongssmbrf}
\end{figure}

\subsection*{Modulation Laser Power}
We have assumed the laser modulator consists of a planar undulator and a TEM00 mode Gaussian laser in an optical enhancement cavity (OEC).
The laser induced modulation voltage in such case is~\cite{Deng2026TLCSSMB}
\begin{equation}\label{eq:TEM00h}
\begin{aligned}
V_{\text{L}}&=\frac{K[JJ]}{\gamma}\sqrt{\frac{2P_{\text{L}}Z_{0}}{\lambda_{\text{L}}}}\frac{\tan^{-1}\left(\frac{L_{u}}{2Z_{R}}\right)}{\sqrt{\frac{L_{u}}{2Z_{R}}}}\sqrt{L_{u}},
\end{aligned}
\end{equation}
where $\gamma$ is the Lorentz factor, $K$ is the undulator parameter of the modulator, $
[JJ]=\left[J_{0}\left(\chi\right)-J_{1}\left(\chi\right)\right],
$
with $\chi=\frac{K^{2}}{4+2K^{2}}$, $P_{\text{L}}$ is the laser power, $Z_{0}=376.73\ \Omega$ is the impedance of free space, $Z_{R}$ is the laser Rayleigh length, and $L_{u}$ is the modulator length.
The optimal Rayleigh length to maximize the modulation is $Z_{R}=0.359L_{u}$. The above formula can be used to evaluate the laser-induced energy modulation strength or the modulation laser power required. We have assumed the average stored laser power in the OEC is about 1 MW, which is close to the state-of-art OEC technology~\cite{Lu2025OEC}. We recognize that the modulator momentum compaction, in addition to making the linear transfer matrix become a thick-lens one, can distort the modulation waveform from the ideal sinusoidal~\cite{Deng2024Springer}, and result also in non-perfect modulation cancellation. This issue can be effectively mitigated by implementing a small section to compensate the modulator momentum compaction inside, or lowering the modulator undulator period number at the expense of a higher laser power required.


\subsection*{Coherent Radiation Power}
As an example, we use a planar undulator as the radiator. Coherent undulator radiation peak power at the fundamental resonance frequency from a transversely-round electron beam is~\cite{Deng2024Springer}
\begin{equation}\label{eq:PR}
P_{\text{peak}}[\text{kW}]=1.183N_{u}\chi[JJ]^{2}FF_{\bot}(S)|b|^{2}\mathcal{D}_{\delta}(\xi)I_{\text{P}}^{2}[\text{A}],
\end{equation}
in which $N_{u}$ is the number of undulator periods, 
$
[JJ]=\left[J_{0}\left(\chi\right)-J_{1}\left(\chi\right)\right],
$
with $\chi=\frac{K^{2}}{4+2K^{2}}$ where $K$ is the undulator parameter of the radiator, and the transverse form factor is
$
FF_{\bot}(S)=\frac{2}{\pi}\left[\tan^{-1}\left(\frac{1}{2S}\right)+S\ln\left(\frac{(2S)^{2}}{(2S)^{2}+1}\right)\right],
$
with $S=\frac{2\pi \sigma^{2}_{\bot}}{\lambda_{\text{R}}L_{u\text{R}}}$ and $L_{u\text{R}}$ is the radiator length, $\sigma_{\bot}=\sqrt{\epsilon_{\bot}\beta_{\bot}}$ the rms transverse electron beam size, $b$ is the bunching factor at the radiation frequency, 
$
\mathcal{D}_{\delta}= \frac{\text{Erf}^{2}(\sqrt{2}\pi\xi)}{8\pi \xi^{2}}
$
with $\xi\equiv N_{u}\sigma_{\delta,\text{eff}}$ is used to quantify the impact of effective energy spread $\sigma_{\delta,\text{eff}}$ at the radiator on coherent radiation power,
and $I_{\text{P}}$ is the peak current. We have assumed an effective $\beta_{\bot}=\frac{L_{uR}}{2}$ in calculating the radiation power given in  Tab.~\ref{tab1}.

\subsection*{BBHG and Weak-Strong Mechanism at Other Wavelengths}


The proposed BBHG method and Weak-Strong mechanism can be scaled to other wavelengths, like RF and THz, for bunch length and energy spread control, ultrashort X-ray pulse and coherent THz generation. The BBHG method can also be applied in a single-pass machine.
Figure~\ref{fig:weakstrongssmbrf} is an example schematic layout of applying the Weak-Strong mechanism at the RF wavelengths, for bunch length and energy spread control in a storage ring, respectively. 




\bmhead{Acknowledgements}

This work is supported by the National Natural Science Foundation of China (NSFC Grant No. 12522512), the National Key Research and Development Program of China (Grant No.~2022YFA1603400), the Beijing Outstanding Young Scientist Program (No. JWZQ20240101006) and the Tsinghua University Initiative Scientific Research Program.

\bmhead{Author contributions}
X.D. conceived the BBHG method and the Weak-Strong SSMB mechanism, and conducted the analysis. All authors contributed to the further crystallization of the ideas. X.D. wrote the manuscript with inputs from all authors. 




\end{document}